---------------------------------------------------------------------------
\date {}
\documentstyle{article}
\begin{document}
\large
\title  {\bf  Rare Decays of $\psi$ and $\Upsilon$}
\author{ K. K. Sharma and R. C. Verma$^*$ \\
\normalsize Centre for Advanced Study in Physics, Department of Physics, \\
\normalsize Panjab University, Chandigarh -160014, India.\\
\normalsize $^{*}$ Department of Physics, Punjabi University, Patiala  - 147 002, India.}

\maketitle
\begin{abstract}
We study two-body weak hadronic decays of $\psi$ and $\Upsilon$
employing the factorization scheme. Branching ratios for $\psi
\rightarrow PP/PV$  and $\Upsilon \rightarrow PP/PV$ decays in the
Cabibbo-angle-enhanced and Cabibbo-angle-suppressed modes are predicted.
\end{abstract}

\newpage
\large
\section{Introduction} A wealth of experimental data on masses,
lifetimes and decay rates of charm and bottom mesons has been collected
[1]. Models based on factorization scheme [2,3], successfully used to
describe weak decays of naked flavor mesons, can also be used to study
weak decays of hidden flavor mesons. Low lying states of quarkonia
systems usually decay through intermediate photons or gluons produced by
the parent $c\bar c$ and $b\bar b$ quark pair annihilation. For
instance, charmonium state $\psi (1S)$ decays predominantly to hadronic
states $(87.7\%)$, whereas its leptonic modes are observed to be around
$6\%$. The same trend follows for the bottomonium states $\Upsilon (1S)$
also. These OZI violating but flavor conserving decays lead to  narrow
widths to $\psi$ and $\Upsilon$ states.
\par In the Standard Model framework, the flavor changing decays of
these states are also possible though these are expected to have rather
low branching ratios. At present, experiments have crossed over to many
million $\psi$-events and the prospectus are good for a 100-fold
increase in these events in the high energy electron-positron collider.
It is expected that some of its rare decay modes may also become
detectable in future. With this possibility in mind, we study two-body
weak decays of $\psi$ in Cabibbo-angle-enhanced and
Cabibbo-angle-suppressed modes. Employing the factorization scheme, we
predict branching ratios for $\psi\rightarrow PP/PV$ decays (~where P
and V represent pseudoscalar meson and vector meson respectively~) and
extend this analysis to $\Upsilon \rightarrow PP/PV$ decays involving
$b\rightarrow c$ transitions.
\section{Two-body Weak Decays of $\psi$}
The structure of the general weak current $\otimes$  current
Hamiltonian is
\begin{equation}  H_{W} = \frac {G_{F}}
{\sqrt 2} V_{ud}V_{cs}^* \{ a_{1} (\bar s c)_{H} (\bar u d)_{H} +
a_{2} (\bar s d)_{H} ( \bar u c)_{H} \} + h.c. ,\end{equation}
for Cabibbo-angle-enhanced mode $(\Delta C=\Delta S= -1)$,
and
$$ H_{W} = \frac {G_{F}}
{\sqrt 2} V_{ud}V_{cd}^* [ a_{1} \{ (\bar u d)_{H} (\bar d
c)_{H} ~-~(\bar u s)_{H} (\bar s c)_{H}\}$$
\begin{equation}
~~~~+~ a_{2} \{ (\bar u c)_{H} ( \bar d d)_{H} ~-~(\bar u
c)_{H} ( \bar s s)_{H} \}] + h.c. ,\end{equation}
for Cabibbo-angle-suppressed mode $(\Delta C = -1, \Delta S = 0)$.
$\bar{q_1}q_2\equiv \bar{q_1}\gamma_\mu (1-\gamma_5)q_2$
represents the color singlet V-A current and $V_{ij}$ denote
Cabibbo-Kobayashi-Maskawa (CKM) mixing matrix elements. The subscript H
implies that ( $\bar {q_1}q_2$) is to be treated as a hadron field
operator. $a's$ are the two undetermined coefficients assigned to the
effective charge current, $a_1$, and the effective neutral current,
$a_2$, parts of the weak Hamiltonian. These parameters can be related to
the QCD coefficients $c_{1,2}$ as
\begin{equation} a_{1,2}=c_{1,2}+\zeta c_{2,1}, \end{equation}
where  $\zeta$ is usually treated as a free parameter, to be fixed by the
experiment. We take
\begin{equation} a_1~~=~~1.26,~~~a_2~~=~~-0.51, \end{equation}
on the basis of $D\rightarrow  K\pi$ decays [4].
\subsection{$\psi \rightarrow PP$ Decays}
The decay rate formula for $\psi \rightarrow PP$ decays is
given by
\begin{equation} \Gamma (\psi \rightarrow PP) \hskip 0.2cm = \hskip
0.2cm \frac {p_{c}^{3}} {24\pi m^2_{\psi}}|A(\psi \rightarrow
PP)|^{2},\end{equation}
where $p_{c}$ is the magnitude of the three momentum of final state
meson in the rest frame of $\psi$ meson and $m_{\psi}$ denote its mass.
Following the procedure adopted by one of us (RCV) with Kamal and
Czarnecki [5] in determination of the weak decay amplitudes of $\psi$
decays, we express the decay amplitude of $\psi \rightarrow PP$ as (upto
the scale $\frac {G_{F}} {\sqrt2}\times CKM factor \times QCD$
coefficient),
\begin{equation} A(\psi \rightarrow PP) ~=~ \langle P \vert J^\mu\vert
0\rangle \langle P \vert J_\mu\vert \psi\rangle, \end{equation}
where $J^\mu$ is the weak V-A current. Matrix elements of the weak current
are given by
\begin{equation} \langle P(k) \vert A_\mu\vert 0\rangle ~=~-\iota ~ f_P
 ~k_\mu, \end{equation}
and
$$ \langle P \vert J_\mu \vert \psi\rangle ~=~\frac {1}{m_{\psi}+m_P}
\epsilon_{\mu\nu\rho\sigma} \epsilon^{\nu}_{\psi} (P_\psi + P_P)^\rho
q^\sigma V(q^2)~ - ~ \iota ~{(m_{\psi}+m_P)} \epsilon^{\mu}_{\psi}
A_1(q^2)$$
$$ ~-~\iota~ \frac {\epsilon_\psi\cdot q}{m_{\psi}+m_P} (P_\psi +
 P_P)^\mu A_2(q^2)  ~+~\iota ~\frac {\epsilon_\psi \cdot q}{q^2}
 (2m_\psi)q^\mu A_3(q^2)$$ \begin{equation}~~~~~~~~ ~-~\iota~ \frac
 {\epsilon_\psi\cdot q}{q^2} (2m_\psi)q^\mu A_0(q^2),\end{equation}
where $f_P$ is the meson decay constant, $\epsilon_\psi$ is the
polarisation vector of $\psi$, $P_\psi$ and $P_P$ are the four-momenta
of $\psi$ and pseudoscalar meson respectively, and $q^\mu
~=~(~P_\psi~-~P_P)^\mu$.  $A_3(q^2)$ is related to $A_1(q^2)$ and
$A_2(q^2)$ as
\begin{equation} A_3(q^2)~=~\frac {(~m_\psi~+~m_P~)}{2m_\psi} A_1(q^2)
~+~\frac {(~m_\psi~-~m_P~)}{2m_\psi} A_2(q^2).\end{equation}
In the Cabibbo-angle-enhanced mode, $\psi$ can decay to $D_s^+\pi^-
or D^0K^0$. To illustrate the procedure, we consider the color enhanced
decay $\psi \rightarrow D_s^+\pi^-$ whose decay amplitude can be
expressed as
\begin{equation} A(\psi \rightarrow D_s^+\pi^-) ~=~\frac {G_{F}} {\sqrt
2} V_{ud}V_{cs}^*~ a_1 ~  \langle \pi^- \vert J^\mu\vert 0\rangle
~\langle D_s^+ \vert J_\mu\vert \psi\rangle, \end{equation}
which gets simplified to
\begin{equation} A(\psi \rightarrow D_s^+\pi^-)~=~\frac {G_{F}} {\sqrt
2} V_{ud}V_{cs}^*~a_1 (2m_\psi )f_\pi (\epsilon_\psi\cdot q)A_0^{\psi
D_s}(m_\pi^2).\end{equation}
Similarly, the factorization amplitude of the decay $\psi \rightarrow
D^0 K^0$ and Cabibbo-angle-suppressed decays of $\psi$ are obtained. We
take the following values for the decay constants (in $GeV$) [6]:
$$f_\pi~=~0.132, ~~~~f_K~=~0.161,
$$ \begin{equation}~f_\eta~=~ 0.131,~~~~ f_{\eta'}~=~0.118.\end{equation}
For the form factors at $q^2 ~=~0$, we use
\begin{equation} A_0^{\psi D}(0)~=~0.61,~   A_0^{\psi D_s}(0)~=~0.66,
\end{equation}
obtained in the earlier work [5] using the BSW model wavefunctions [7]
at $\omega ~=~ 0.5$ GeV. The oscillator parameter $\omega$ is the
measure of the average transverse momentum of the quark in the meson
[7]. We use the following basis for $\eta-\eta'$ mixing:
\begin{equation} \eta  = \frac {1}{\sqrt 2} (u \bar u + d \bar d) \sin
\phi_{p} - (s \bar s) \cos \phi_{p}, \end{equation}
\begin{equation} \eta^{\prime}  = \frac {1}{\sqrt 2} (u \bar u + d \bar
d) \cos \phi_{p} + (s \bar s) \sin \phi_{p},\end{equation}
where $ \phi_{p}~ =~ \theta_{ideal}~  - ~\theta _{physical}$.
Using the decay rate formula (5), we compute branching ratios of
various $\psi \rightarrow PP$ decays which are listed in Table-1.
Among the Cabibbo-angle-enhanced decays, we find that the dominant decay
is $\psi\rightarrow D_s^+\pi^-$, having the branching ratio
\begin{equation} B(\psi\rightarrow D_s^+\pi^-)~=~(0.87\times
10^{-7})\%,\end{equation}
and the next in order is $\psi\rightarrow D^0K^0$, whose branching ratio
is
\begin{equation} B(\psi\rightarrow D^0K^0)~=~(0.28\times
10^{-7})\%.\end{equation}
We hope that these branching ratios would lie in the detectable range.
\subsection{$\psi \rightarrow PV~ Decays$}
Similar to $\psi \rightarrow PP$ decays, the weak decay amplitude for
$\psi \rightarrow PV$ decays can be expressed as product of the matrix
elements of the weak currents,
\begin{equation} A(\psi \rightarrow PV) ~=~ \langle V \vert J^\mu\vert
0\rangle \langle P \vert J_\mu\vert \psi\rangle,  \end{equation}
where the vector meson is generated out of the
vacuum, and the corresponding matrix element is given by
\begin{equation} \langle V(k) \vert V_\mu\vert 0\rangle
~=~\epsilon_\mu^*m_Vf_V. \end{equation}
Using the matrix elements given in (8), and (9), we obtain
$$ A(\psi \rightarrow PV)=[\frac {2m_Vf_V }{m_{\psi}+m_P}
\epsilon_{\mu\nu\rho\sigma} \epsilon_1^{*\mu}
\epsilon_2^{*\nu}P_\psi^\rho P_P^\rho V(q^2) $$
$$+ \iota(m_Vf_V)\{
\epsilon_1^{*}.\epsilon_2^*(m_{\psi}+m_P)A_1(q^2)$$
\begin{equation}~~~~~~~~~~~~~~ ~-~\epsilon_1^*.(P_\psi -
P_P)\epsilon_2^*.\frac {P_\psi + P_P}{m_{\psi}+m_P} A_2(q^2) \}],
\end{equation}
which yields the following decay rate formula:
$$ \Gamma (\psi \rightarrow PV) \hskip 0.2cm = \hskip 0.2cm
 (nonkinematic~ factor)^2 \frac {p_c} {24\pi~m_\psi^2}
 (~m_V~f_V~)^2(~m_\psi~+~m_P~)^2$$ \begin{equation}~~~\times \{~
 \alpha~|V(q^2)|^2~+~\beta~|A_1(q^2)|^2 ~+~\gamma~|A_2(q^2)|^2 ~+~\delta
 ~ Re [~A^*_1(q^2)~ \times ~A_2(q^2)]\},\end{equation}
where
\begin{equation} \alpha ~=~ \frac {8~m_P^2~p_c^2}{(~m_P~+~m_\psi ~)^4},
\end{equation}
\begin{equation} \beta~=~ 2~+~\{\frac
{m_\psi^2~-~m_P^2~-~m_V^2}{2~m_\psi~m_V}\}^2 ,\end{equation}
 \begin{equation} \gamma ~=~ \frac {4~m_P^4~p_c^4}{
m_\psi^2~m_V^2~(~m_P~+~m_\psi ~)^4}, \end{equation}
and
\begin{equation} \delta ~=~ \frac
{2~(m_\psi^2~-~m_P^2~-~m_V^2)}{(~m_P~+~m_\psi ~)^2} \frac
{~m_P^2~p_c^2}{ m_\psi^2~m_V^2~}, \end{equation}
Nonkinematic factor is the product of scale factor ($\frac
{G_{F}} {\sqrt2}V_{ud} Vcs^*$) and the appropriate QCD coefficient $a_1$
or $a_2$. The terms corresponding to $A_1(q^2)$, $V(q^2)$, and
$A_2(q^2)$ represent S, P, and D partial waves in the final state. We
have computed the numerical coefficients (~$\alpha$, $\beta$, $\gamma$,
$\delta$~) for various decays which are given in Table-2. We observe
that the numerical coefficient of $A_1(q^2)$ is the largest, thus the
retention of S-wave only would appear to be an excellent approximation.
For the sake of simplicity, we have retained only this term in our
analysis. Following values of the vector meson decay constants (in
$GeV$) [6] are used in our analysis:
\begin{equation} f_\rho~=~0.216,
~f_{K^*}~=~0.221,~f_\omega~=0.195,~f_\phi~=~0.237,
f_{D^*}~=0.250,\end{equation}
and the form factors [5]
\begin{equation} A_1^{\psi D}(0)~=~0.68,~ ~~  A_1^{\psi D_s}(0)~=~0.78,
\end{equation}
at $\omega ~=~ 0.5$ GeV. Using the decay rate formula (21), we obtain
the branching ratios of $\psi \rightarrow PV$ decays in
Cabibbo-angle-enhanced and Cabibbo-angle-suppressed modes which are
given in Table-3. For the color enhanced decay of the
Cabibbo-angle-enhanced mode, we calculate \begin{equation}
B(\psi\rightarrow D_s^+\rho^- )~=~ (0.36\times 10^{-6}),\%\end{equation}
which is higher than the branching ratio of $\psi \rightarrow
D_s^+\pi^-$. Our analysis yields
\begin{equation} \frac {B(\psi\rightarrow D_s^+\rho^-
)}{B(\psi\rightarrow D_s^+\pi^- )}~=~4.2, \end{equation}
and therefore $\psi\rightarrow D_s^+\rho^-$ can be expected to be
measured soon.
\section{Two-body Weak Decays of $\Upsilon$}
In this section, we extend our analysis to $ \Upsilon \rightarrow
PP / PV $ decays.
\subsection{$ \Upsilon \rightarrow PP$ ~Decays}
The  effective weak Hamiltonian generating the dominant
b quark decays involving $ b \rightarrow c $ transition
is given by
$$ H^{\Delta b = 1 }_W = \frac {G_F} {\sqrt 2} \{   V_{cb} V^{*}_{ud} [
      a_1 (\bar{c}b) (\bar{d}u) + a_2 (\bar{d}b) (\bar{c}u) ] $$
\begin{equation}~~~+ V_{cb} V^{*}_{cs} [ a_1 (\bar{c}b) (\bar{s}c) + a_2
      (\bar{s}b) (\bar{c}c) ] \}  + h.c.,  \end{equation}
for the CKM favored mode and
$$  H^{\Delta b = 1 }_W = \frac {G_F} { \sqrt{2}} \{  V_{cb} V^{*}_{us}
      [ a_1 (\bar{c}b) (\bar{s}u) + a_2 (\bar{s}b) (\bar{c}u) ]$$
\begin{equation} ~~~      +  V_{cb} V^{*}_{cd} [ a_1 (\bar{c}b)
    (\bar{d}c) + a_2 (\bar{d}b) (\bar{c}c) ]\} + h.c.,\end{equation}
for the CKM singly suppressed mode. In our analysis we use
\begin{equation} a_1~= ~1.03,~~ a_2~ =~0.23, \end{equation} as
guided by $ B \rightarrow  PP / PV$ data [8]. Similar to $\psi
\rightarrow PP$ decays, the  factorization scheme expresses weak
decay amplitudes as  a product of matrix elements of the weak
currents ( upto the scale    $ \frac {G_F} {\sqrt{2}} \times CKM
factor \times QCD factor )$ as :
\begin{equation} A ( \Upsilon \rightarrow P P)~  = ~ < P | J^\mu | 0 >
< P | J_\mu | \Upsilon >~.  \end{equation}
For instance the decay amplitude for the color enhanced mode
$ \Upsilon \rightarrow B_c^+ \pi^- $ of the CKM-favored decays
is given by
\begin{equation} A ( \Upsilon \rightarrow B_c^+ \pi^-)~ = ~\frac
 {G_F}{\sqrt{2}} \times V_{cb} V^{*}_{ud} \times a_1 f_\pi ( 2 m_
 \Upsilon ) A_0^{\Upsilon \rightarrow B_c}( m^{2}_\pi).\end{equation}
We take (in $GeV$)
\begin{equation} f_D~ =~ 0.240,~ f_{D_S} ~ = ~0.280, ~f_{\eta_c} ~ =
   ~0.393,\end{equation}
values [6] in our calculations. Generally the form factors for the weak
transitions between heavy mesons are calculated in the quark model
framework using meson wave functions. However, in the past few years the
discovery of new flavor and spin symmetries has simplified the heavy
flavor physics [9], and it has now become possible to calculate these
form factors from certain mass factors involving the Isgur-Wise
function. In the framework of heavy quark effective theory (HQET), these
can be expressed as
\begin{equation} A^{\Upsilon\rightarrow B_c} _0(q^2)~=~ \frac {
 m_\Upsilon + m_{B_c}}{2\sqrt {( m_{B_c}m_\Upsilon )}} \xi
 (\omega),\end{equation}
with
\begin{equation} \omega ~=~ v_\Upsilon\cdot v_{B_c}~=~\frac
{m_{\Upsilon}^2~+~m_{B_c}^2~-~m_{\pi}^2}{2~m_{\Upsilon}~m_{B_c}},\end{equation}
where the Isgur-Wise function $\xi$ is normalized to unity at kinetic
point $v_{\Upsilon}\cdot v_{B_c} ~=~1$. As $\omega~\approx~1.08$ for
$\Upsilon \rightarrow B_c^+ \pi^-$ decay, we have ignored the $\omega$
dependence of the Isgur-Wise function $\xi$ and calculate
\begin{equation} A_0^{\Upsilon \rightarrow B_c}(q^2) \approx
0.98 .\end{equation}
This in turn yields
\begin{equation} B( \Upsilon \rightarrow B_c^+ \pi^-)~ = 0.33\times
   10^{-8}\% . \end{equation}
The decay amplitudes for other CKM-favored  and CKM-suppressed decay
modes are obtained similarly. Branching ratios for various $ \Upsilon
\rightarrow PP $ decays are given in Table 4. We find that the dominant
decays are $\Upsilon\rightarrow B_c^+D_s^-$ and  $\Upsilon\rightarrow
B_c^+\pi^-$.
\subsection{$\Upsilon \rightarrow PV$ ~Decays}
The decay rate formula for such decays has been discussed in the section
3. Here also for comparison of  the contributions of the various form
factors involved,  we have calculated the numerical coefficients (~$
\alpha,~ \beta,~ \gamma,~ \delta~)$ for various $ \Upsilon \rightarrow
PV $  decays, which are given in Table 5. Like $ \psi \rightarrow PV $
decays, here also numerical coefficient ($\alpha$) of  $A_1( q^2)$ is
found to be the largest. Various form factors appearing in
$A(\Upsilon\rightarrow PV)$ are mutually related by HQET [9],
$$ \frac { m_\Upsilon + m_{B_c}}{2\sqrt {( m_{B_c}m_\Upsilon )}} \xi
  (\omega) ~=~V(q^2)~=~A_0(q^2)~=~A_2(q^2)$$
\begin{equation}~~~~~~ =~\{~1~- ~ \frac{q^2}{(m_{\Upsilon}
  ~+~m_{B_c}~)^2} \}^{-1} A_1(q^2),\end{equation}
 \begin{equation}q^2~=~m_{\Upsilon}^2~+~m_{B_c}^2~-~2~m_{\Upsilon}~m_{B_c}~v_{\Upsilon}\cdot
 v_{B_c}.\end{equation}
Following the procedure used for $\psi \rightarrow PV$ decays, we
determine the decay amplitudes for various  $ \Upsilon \rightarrow PV$
decays in the CKM-favored and  CKM-suppressed modes. In addition to the
meson decay constants given in (26), two more decay constants ( in $
GeV$ ) [6]
\begin{equation} f_\psi~=~0.405,~~f_{D_s^*}~=~0.271, \end{equation}
are used here. Branching ratios for these decays are given in Table 6.
Here also, we observe that the dominant mode is $B(\Upsilon \rightarrow
B_c^+D_s^{*-}) ~=~ (2.57\times 10^{-8})\%$, which is higher than
$B(\Upsilon \rightarrow B_c^+\pi^-)$  by a factor of 7.9.

\vskip 1.0 cm
\noindent {\bf Acknowledgements}\\ 
One of the authors (RCV) gratefully acknowledges the financial support from
the Department of Science and Technology, New Delhi, India.

\newpage
\begin{table} \begin{center} \caption {Branching Ratios
of $\psi\rightarrow PP$ decays}  \vskip 0.3 cm
\begin{tabular}{|c|c|c|} 
\hline 
Mode & Decay  & $ Br.(\times 10^{-8} \%)$ \\
\hline 
$\Delta C = \Delta S = + 1 $ & & \\
 & $ \psi \rightarrow D_s^+ \pi^{-}$ & $8.74$\\
 & $\psi \rightarrow D^{0}K^0$ &  $2.80$ \\
$\Delta C = +1, \Delta S = 0 $ & & \\
 & $ \psi \rightarrow D_s^+K^{-}$ & $0.55$ \\
 & $ \psi \rightarrow D^+ \pi^{-}$ & $0.55$ \\
 & $ \psi \rightarrow D^0 \eta $ & $0.016$ \\ 
 & $ \psi \rightarrow D^0 \eta' $ & $0.003$ \\ 
 & $ \psi \rightarrow D^0 \pi^0 $ & $0.055$ \\ 
\hline
\end{tabular} \end{center} \end{table}

\newpage
\begin{table} \begin{center} \caption {Numerical Coefficients of the
Form Factors for $\psi \rightarrow PV$ Decays}
\vskip 0.3 cm
\begin{tabular}{|c|c|c|c|c|} \hline Decay  &  $\alpha $
&$\beta$ & $\gamma $ & $\delta$ \\ \hline \hline
$ \psi \rightarrow D_s^+ \rho^{-}$& $0.0210$ & $3.755$ & $0.0032$ & $0.150$\\
$ \psi \rightarrow D^{0}K^{*0}$  &  $0.0204$ & $3.553$ & $0.0020$ & $0.113$\\
 &&&& \\
$ \psi \rightarrow D_s^+ K^{*-}$ &  $0.0146$ & $3.390$ & $0.0011$ & $0.080$\\
$ \psi \rightarrow D^+ \rho^-$   &  $0.0264$ & $3.974$ & $0.0047$ & $0.192$\\
$ \psi \rightarrow D^0 \rho^{0}$ &  $0.0267$ & $3.984$ & $0.0048$ & $0.194$\\
$ \psi \rightarrow D^0 \omega $  &  $0.0261$ & $3.928$ & $0.0044$ & $0.184$\\
$ \psi \rightarrow D^0 \phi $    &  $0.0135$ & $3.284$ & $0.0007$ & $0.060$\\
\hline
\end{tabular}
 \end{center} \end{table}

\newpage
\begin{table} \begin{center} \caption {Branching Ratios
of $\psi\rightarrow PV$ decays}  \vskip 0.3 cm
\begin{tabular}{|c|c|c|} 
\hline 
Mode & Decay  & $ Br.(\times 10^{-8} \%)$  \\
\hline  
$\Delta C = \Delta S = +1 $ & & \\
 & $\psi \rightarrow D_s^+ \rho^{-}$ & $36.30$ \\
 & $ \psi \rightarrow D^{0}K^{*0}$ &  $10.27$\\
$\Delta C = +1, \Delta S = 0 $  & & \\
 & $ \psi \rightarrow D_s^+ K^{*-}$ & $2.12$\\
 & $ \psi \rightarrow D^+ \rho^-$ & $2.20$\\
 & $ \psi \rightarrow D^0 \rho^{0}$ & $0.22$\\
 & $ \psi \rightarrow D^0 \omega $ & $0.18$\\
 & $ \psi \rightarrow D^0 \phi $ & $0.65$\\
\hline
\end{tabular}
 \end{center} \end{table}

\newpage
\begin{table} \begin{center} \caption {Branching Ratios
of $\Upsilon \rightarrow PP$ decays}  \vskip 0.3 cm
\begin{tabular}{|c|c|c|} 
\hline 
Mode & Decay  & $ Br.(\times 10^{-8} \%)$ \\ 
\hline 
$\Delta b = 1, \Delta C = 1, \Delta S = 0$  & & \\
& $ \Upsilon \rightarrow B_c^+ \pi^{-}$ & $0.33$ \\
 & $ \Upsilon \rightarrow B^0D^{0}$ &  $0.10$ \\
$\Delta b = 1, \Delta C = 0, \Delta S = -1$ & & \\
& $ \Upsilon \rightarrow B_c^+D_s^-$ & $0.76$ \\
 & $ \Upsilon \rightarrow B_s^0\eta_c$ & $0.17$ \\
$\Delta b = 1, \Delta C = 1, \Delta S = -1 $& & \\ 
& $ \Upsilon \rightarrow B_c^+K^-$ & $0.024$ \\
 & $ \Upsilon \rightarrow B_s^0D^0$ & $0.005$ \\ 
$\Delta b = 1, \Delta C = 0, \Delta S = 0$ & & \\
& $ \Upsilon \rightarrow B_c^+D^-$ & $0.031$ \\
 & $ \Upsilon \rightarrow B^0\eta_c$ & $0.010$\\
\hline
\end{tabular}
 \end{center} \end{table}

\newpage
\begin{table} \begin{center} \caption {Numerical Coefficients of the
Form Factors for $\Upsilon \rightarrow PV$ Decays}
\vskip 0.3 cm
\begin{tabular}{|c|c|c|c|c|} \hline 
Decay  &  $\alpha $
 & $\beta$ & $\gamma $ & $\delta$ \\ 
\hline
$ \Upsilon \rightarrow B_c^+ \rho^{-}$ & $0.0319$ & $13.414$& $0.0759$ & $1.862$ \\
$ \Upsilon \rightarrow B^0 D^{*0}$     & $0.0379$ & $5.000$ & $0.0117$ & $0.375$ \\ &&&&\\
$ \Upsilon \rightarrow B_c^+ D_s^{*-}$ & $0.0175$ & $3.758$ & $0.0030$ & $0.146$ \\
$ \Upsilon \rightarrow B_s^0 \psi$     & $0.0199$ & $3.435$ & $0.0014$ & $0.090$ \\ &&&& \\
$ \Upsilon \rightarrow B_c^+ K^{*-}$   & $0.0311$ & $10.550$& $0.0537$ & $1.355$ \\
$ \Upsilon \rightarrow B_s^0 D^{*0}$   & $0.0365$ & $4.898$ & $0.0112$ & $0.360$ \\ &&&& \\
$ \Upsilon \rightarrow B_c^+ D^{*-}$   & $0.0191$ & $3.911$ & $0.0040$ & $0.174$ \\
$ \Upsilon \rightarrow B^0 \psi$       & $0.0216$ & $3.477$ & $0.0016$ & $0.097$ \\
\hline
\end{tabular}
 \end{center} \end{table}

\newpage
\begin{table} \begin{center} \caption {Branching Ratios
of $\Upsilon \rightarrow PV$ decays}  \vskip 0.3 cm
\begin{tabular}{|c|c|c|} 
\hline 
Mode & Decay  & $ Br.(\times 10^{-8} \%)$  \\  
\hline
$\Delta b = 1, \Delta C = 1, \Delta S = 0$ & & \\
& $ \Upsilon \rightarrow B_c^+ \rho^{-}$ & $0.88$ \\
 & $ \Upsilon \rightarrow B^0 D^{*0}$ & $0.19$ \\
$\Delta b = 1, \Delta C = 0, \Delta S = -1$ & & \\ 
& $ \Upsilon \rightarrow B_c^+ D_s^{*-}$ & $2.57$ \\
 & $ \Upsilon \rightarrow B_s^0 \psi$ & $0.93$ \\
$\Delta b = 1, \Delta C = 1, \Delta S = -1$ & & \\
& $ \Upsilon \rightarrow B_c^+ K^{*-}$ & $0.050$ \\
 & $ \Upsilon \rightarrow B_s^0 D^{*0}$ & $0.009$ \\
$\Delta b = 1, \Delta C = 0, \Delta S = 0$ & & \\
& $ \Upsilon \rightarrow B_c^+ D^{*-}$ & $0.11$\\
 & $ \Upsilon \rightarrow B^0 \psi$ & $0.053$ \\
\hline
\end{tabular}
 \end{center} \end{table}
\end{document}